\documentclass[twocolumn,resetfootnote]{aastex7}

\usepackage[utf8]{inputenc}
\usepackage{tipa}
\usepackage{combelow}
\usepackage{savesym}
\savesymbol{tablenum}
\usepackage{siunitx}
\restoresymbol{SIX}{tablenum}
\usepackage{breqn}
\usepackage{tabularx}
\usepackage{placeins}

\begin{document}

\title{Orbital characterization of a newly discovered small satellite around Quaoar}

\author[orcid=0000-0002-1788-870X,sname='Proudfoot']{Benjamin Proudfoot}
\affiliation{Florida Space Institute, University of Central Florida, 12354 Research Parkway, Orlando, FL 32826, USA}
\email[show]{benp175@gmail.com}

\author[orcid=0009-0006-2600-196X]{Richard Nolthenius}
\affiliation{Cabrillo College Astronomy Department, 6500 Soquel Drive, Aptos, CA, 95003, USA}
\email[]{}

\author[orcid=0000-0002-6117-0164]{Bryan J. Holler}
\affiliation{Space Telescope Science Institute, Steven Muller Building, 3700 San Martin Drive Baltimore, MD 21218 Baltimore, MD, USA}
\email{}

\author[orcid=0000-0002-2414-1460]{Ana Carolina de Souza-Feliciano}
\affiliation{Florida Space Institute, University of Central Florida, 12354 Research Parkway, Orlando, FL 32826, USA}
\email{}

\author[orcid=0000-0002-6085-3182,sname='Rommel']{Flavia L. Rommel}
\affiliation{Florida Space Institute, University of Central Florida, 12354 Research Parkway, Orlando, FL 32826, USA}
\email{}

\author[0009-0004-7149-5212]{Cameron Collyer}
\affiliation{Florida Space Institute, University of Central Florida, 12354 Research Parkway, Orlando, FL 32826, USA}
\email{}

\author[orcid=0000-0002-8296-6540, sname='Grundy']{Will M. Grundy} 
\affiliation{Lowell Observatory, 1400 W Mars Hill Rd, Flagstaff, AZ 86001, USA}
\affiliation{Northern Arizona University, Department of Astronomy \& Planetary Science, PO Box 6010, Flagstaff, AZ 86011, USA}
\email{}

\author[0000-0003-2132-7769]{Estela Fern\'{a}ndez-Valenzuela}
\affiliation{Florida Space Institute, University of Central Florida, 12354 Research Parkway, Orlando, FL 32826, USA}
\email{}


\begin{abstract}

Recent observations of a stellar occultation have revealed the presence of a previously undiscovered small satellite around Quaoar. Orbiting near Quaoar's unusual ring system, this new satellite has the potential to provide significant insights into the formation and evolution of Quaoar and its ring system. In this letter, we characterize the orbit of this newly discovered satellite, finding that it is likely on a $3.6^{+0.5}_{-0.3}$-day orbit, plausibly placing it near a 5:3 mean motion resonance with Quaoar's outermost known ring. Examining the possibility of observing the newly discovered satellite with further stellar occultations, we estimate that $\sim$hundreds of observing stations are required for recovery, since phase information about its orbit was rapidly lost after the lone detection. We also attempted to recover the satellite in JWST NIRCam imaging of Quaoar, but find no convincing detection. This non-detection is limited by the accuracy of the available NIRCam PSF models, as well as the satellite's extreme faintness and close-in orbital separation. Therefore, current-generation telescopes will likely struggle to directly image this new satellite, but near-future 30-meter-class telescopes should prove capable of detecting it. Discovery of such a satellite provides evidence that the rings around Quaoar may have been part of an initially broad collisional disk that has evolved considerably since its formation. To further explore this hypothesis, we encourage follow-up observations of the rings and satellites with stellar occultations and direct imaging, as well as updated hydrodynamical, collisional, and tidal modeling of the system.

\end{abstract}


\section{Introduction} 
\label{sec:intro}

In the past decade, stellar occultations have revealed the presence of rings around small bodies \citep{braga2014ring,ortiz2017size,morgado2023dense,ortiz2023changing}. Among these small body ring systems, the rings around Quaoar are perhaps the most mysterious. The two rings so-far discovered are well-outside the Roche limit and are inhomogeneous \citep{morgado2023dense,pereira2023two,proudfoot2025constraints}. Quaoar's outer ring---named Q1R---appears to be at least partially confined by mean motion resonances (MMRs) with Quaoar's moon Weywot, as well as spin-orbit resonances (SORs) with Quaoar's triaxial shape \citep{morgado2023dense}. The inner ring---Q2R---is seemingly less dense and its confinement is more uncertain. Understanding the formation and long-term stability of these rings provides an opportunity to decipher the formation of large terrestrial bodies in the early epochs of our solar system.

Recently, during a stellar occultation, simultaneous dropouts from two telescopes revealed the presence of a previously undiscovered satellite or dense ring around Quaoar \citep{Nolthenius_2025}. The duration of the dropout implied a minimum diameter/width of 30 km. Here, we argue that a ring (or ring arc) is strongly disfavored for three main reasons. First, the optical depth ($\tau$) required to match the near total flux drop observed \textit{strongly} disfavors the ring hypothesis. Although all of Quaoar's known rings are outside of the Roche limit, they are thought to be stabilized by the cold temperatures of the Kuiper belt \citep[$\sim$40 K,][]{morgado2023dense}, which makes particle collisions more elastic. This mechanism, however, is only effective at $\tau<0.25$, while the recent occultation implies $\tau>6$ in the ring scenario \citep{Nolthenius_2025}, making a ring (or ring arc) prone to rapid accretion. Since accretion progresses on decadal timescales \citep{Kokubo2000,Takeda200}, such a ring would have to be extremely young ($<$decades old). 

Second, at the radial location of the detection, rings are likely to be unstable. Rings subject to strong perturbations from satellites and the non-spherical shape of Quaoar are prone to collisional destruction, especially if the perturbations are of a similar order of magnitude in strength \citep{marzari2020ring}. At the putative ring's radius, the strength of Quaoar's and Weywot's orbital perturbations are nearly equal, making the collisional environment extremely active; therefore, a ring would be warped, have differential precession, and undergo significant viscous spreading. In contrast, a satellite would have no such stability issues.

Last, as pointed out by \citet{Nolthenius_2025}, previous high-precision occultations have shown no signal of a ring at a similar location. Although this detection could represent a dense arc, ring arcs tend to be embedded in a less dense, continuous ring. No such rings have been discovered, despite extensive searches \citep{morgado2023dense,pereira2023two,proudfoot2025constraints}.

Based on the random orientation of the occultation chord during the serendipitous discovery, probabilistic arguments suggest that the newly discovered satellite has a \added{most probable} diameter of $\sim$38 km, \added{although any value greater than 30 km is possible. Given the numerous occultations of Quaoar in the past two decades \citep[e.g.,][]{braga2013size,morgado2023dense,pereira2023two}, the late timing of the recent serendipitous discovery favors a small size. Thermal observations appear to show no excess thermal emission \citep{kiss2024visible}, also favoring a small satellite. Throughout this work, we use an upper limit of 100 km on the diameter to bracket the likely size range of the newly discovered object.} 

\added{Assuming the satellite is $\sim$38 km, the satellite is estimated to have} a $V$ magnitude of $\sim$28, likely making this the faintest satellite ever found around a trans-Neptunian object (TNO). 
In this letter, we explore this detection in greater detail, with a focus on understanding the orbit and recoverability of the satellite. 

\section{Orbit Characterization}
\label{sec:results}

Although only detected at a single epoch, even one observation can provide valuable constraints on the satellite's orbit using a few realistic assumptions. 

Assuming a circular orbit, the minimum semi-major axis is given by its projected distance from Quaoar. The detection had an on-sky offset from Quaoar of ($\Delta \alpha,\Delta \delta$) = $(-0.167\pm0.005, -0.062\pm0.004)$ based on the NIMAv19 ephemeris \citep{desmars2015orbit}\footnote{\url{https://lesia.obspm.fr/lucky-star/obj.php?p=1177}}, which corresponds to a projected on-sky distance of 5380$\pm$150 km. We take this as a plausible lower bound on the semi-major axis.

One way to estimate the maximum semi-major axis is to impose dynamical stability requirements for interactions with Weywot. Looking to planetary stability literature, a pair of orbiting planets (akin to a pair of orbiting satellites) are said to be unstable if their orbits are within $2\sqrt{3}$ mutual Hill radii of each other \citep{birn1973stability}. 
Assuming initially circular orbits, a density of 1000 kg m$^{-3}$ for Weywot, and a neglibly small mass for the new satellite, this predicts $a\lesssim9500$ km. 

Now looking to the eccentricity, tidal damping arguments suggest a low eccentricity. Recent studies have shown that Weywot's eccentricity is low \citep{proudfoot2024bpm2,braga2025investigating}, implying effective eccentricity damping. On the other hand, in the other TNO satellite system with two moons---the Haumea system---the small inner moon, Namaka, has its eccentricity excited ($e\sim0.2$) by mutual interactions with the larger external satellite, Hi'iaka. Unfortunately, without another detection, the satellite's eccentricity is unconstrained, so we proceed under the assumption of a circular orbit, although we revisit this assumption later.



\begin{figure}[b]
    \centering
    \includegraphics[width=\columnwidth]{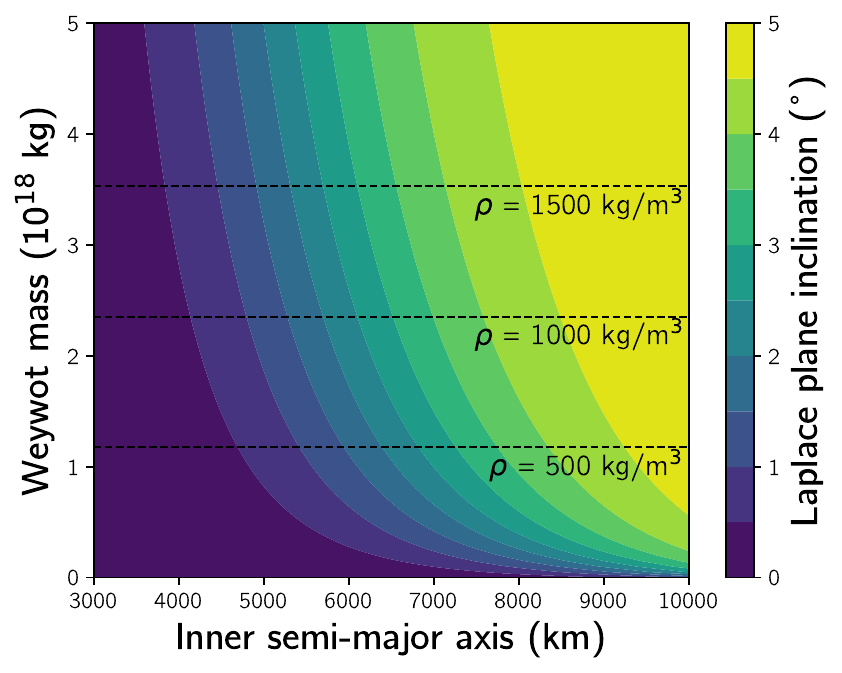}
    \caption{The inclination of the local Laplace plane as a function of the inner satellite's semi-major axis and Weywot's mass. Dotted lines show Weywot's expected mass from different densities based on Weywot's diameter of $\sim$165 km (Fern\'{a}ndez-Valenzuela et al., in prep.). }
    \label{fig:laplace}
\end{figure}


Whereas one may be tempted to assume that the newly discovered satellite's inclination is $\sim$0$\degr$ (i.e., equatorial), analytical theory shows this is very unlikely. Perturbations to the orbit will primarily come from two sources. First, Quaoar's dynamical oblateness, commonly parameterized as $J_2$, and second, Weywot. Both of these perturbations will result in precession of a body's apses and nodes. Since Weywot is $\sim$5$\degr$ inclined from Quaoar's equatorial plane \citep[][Proudfoot et al., submitted]{braga2025investigating}, even if the newly discovered satellite is on an equatorial orbit, it will begin to precess around Weywot's orbital plane, creating a non-zero inclination. The competition between $J_2$ and Weywot perturbations is minimized at the Laplace plane, defined as the mean plane around which a body's orbit pole will precess. The inclination of the Laplace plane from Quaoar's equatorial plane ($\beta$) is given by:
\begin{equation}
    \dot{\Omega}_{w} \sin{(i_w - \beta)} = \dot{\Omega}_{J_2} \sin{\beta}
\end{equation}
\noindent where $\dot{\Omega}_{w}$ and $\dot{\Omega}_{J_2}$ are the precession rates caused by Weywot and Quaoar's $J_2$, respectively, and $i_w$ is Weywot's inclination with respect to Quaoar's equator \citep{burns1979thickness}. Near Quaoar, the $J_2$ precession dominates and the Laplace plane inclination is small. Conversely, near Weywot, the Laplace plane is close to Weywot's orbital plane. 

If the new satellite's orbit has had its inclination effectively damped (i.e., not excited), it should lie in (or near) the Laplace plane. In Figure \ref{fig:laplace}, we show the Laplace plane inclination as a function of the satellite's semi-major axis and Weywot's mass, which has yet to be measured. We assume $J_2=0.018$, which was recently measured in Proudfoot et al. (submitted). Between the minimum and maximum $a$, the Laplace plane can be between $\sim$0.5-4.5$\degr$, with a typical inclination of $\sim$2$\degr$.

\begin{figure}
    \centering
    \includegraphics[width=\columnwidth]{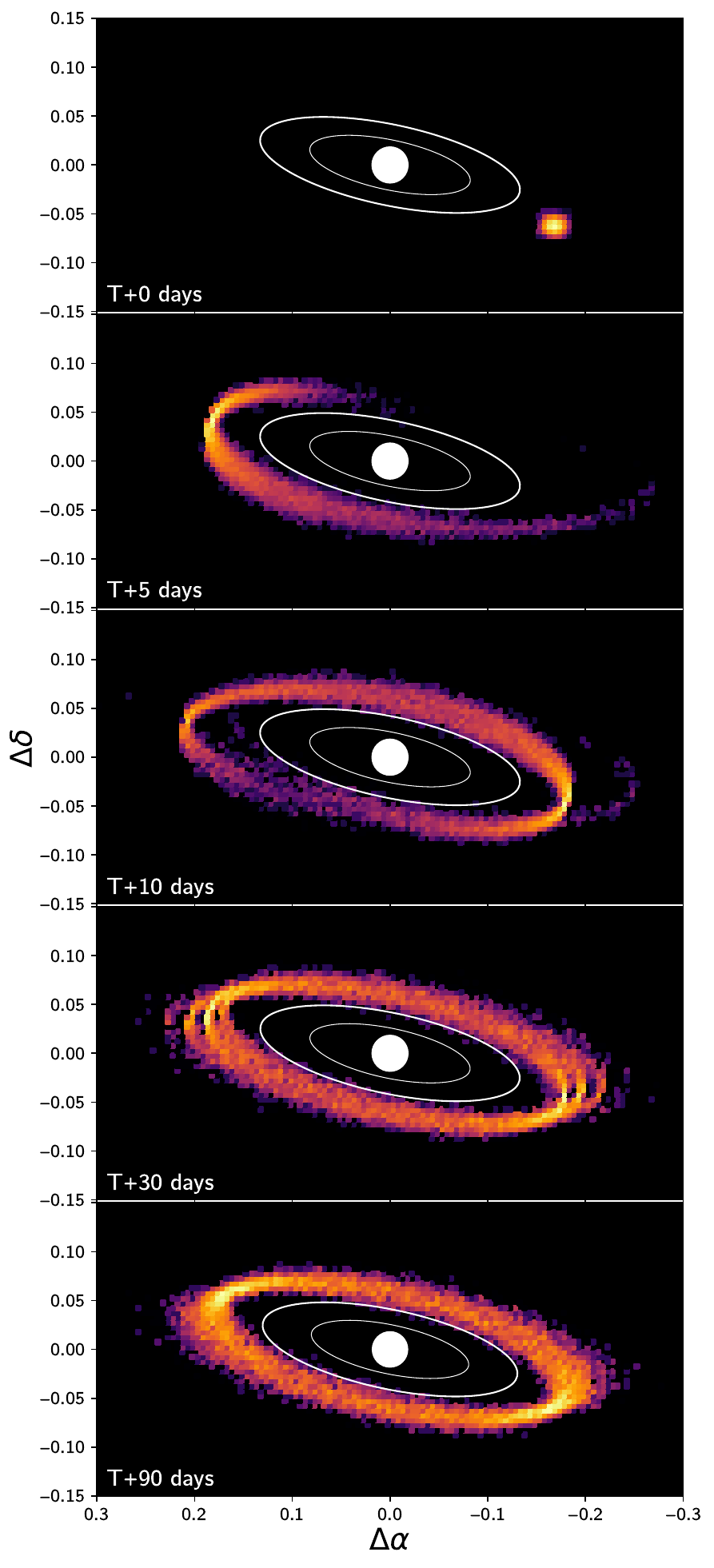}
    \caption{The evolution of the new satellite's on-sky probability density after discovery with high densities shown by brighter color. The 2-d histogram is log-scaled and normalized at each individual epoch to better show the evolution of the distribution. After detection, Keplerian shear rapidly broadens the probability distribution until the distribution is roughly uniform in phase after 1-2 months.}
    \label{fig:evolution}
\end{figure}

\begin{figure}
    \centering
    \includegraphics[width=\columnwidth]{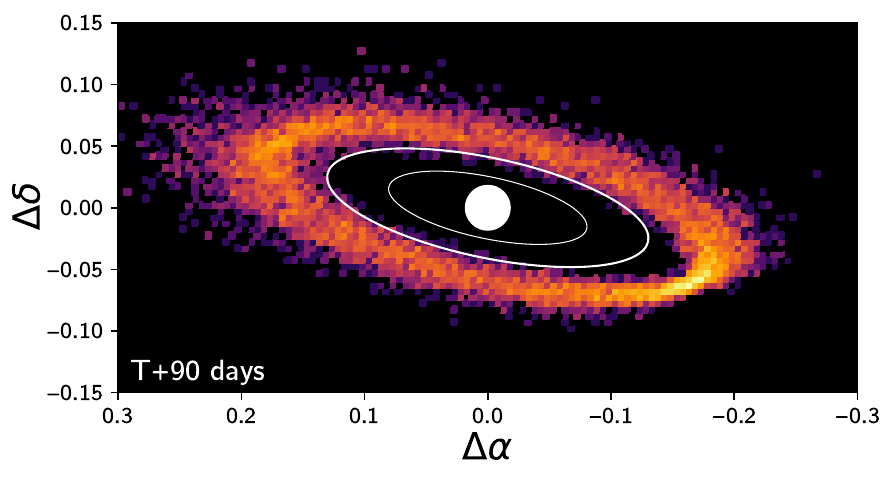}
    \caption{Similar to Figure \ref{fig:evolution}, but for an orbit fit where eccentricity was allowed to be a free parameter. Aside from a slight expansion of the probability distribution away from the intial detection, the main effect is to eliminate the pinch at the top left of the probability distribution.}
    \label{fig:ecc}
\end{figure}

\subsection{Orbit fitting}

With some constraints, the new satellite's orbit can be modeled using a Bayesian approach that incorporates our knowledge/assumptions about the orbit using prior probability distributions. \texttt{MultiMoon}, a Bayesian orbit fitter designed for TNOs, provides a good platform to do this \citep{ragozzine2024beyond}. The orbit model we use is a circular Keplerian orbit, which provides an excellent approximation of the orbit over short timeframes. We place a prior on Quaoar's mass as the measurement provided from analysis of Weywot's orbit, $M_Q=(1212\pm5)\times10^{18}$ kg (Proudfoot et al., submitted). Likewise, we place a prior constraining the inclination (w.r.t Quaoar's equatorial plane) to be 2 $\pm$ 2$\degr$, which roughly approximates the inclined Laplace plane. Although this scheme does not provide a unique solution for the orbit, it can be used to understand the range of possible orbits it might populate. 

The results of this modeling demonstrate that the new satellite's likely semi-major axis is $a = 5838^{+512}_{-326}$ km, corresponding to orbit periods of $P_{orb} = 3.6^{+0.5}_{-0.3}$ days. Using samples of the orbit from the \texttt{MultiMoon} fits, we can also visualize the probability density of the satellite's on-sky position after detection in Figure \ref{fig:evolution}. Soon after the initial discovery, its location was relatively tightly constrained. However, after just a few orbit periods, the cloud of possible positions shears out considerably. After $\sim$30 days all phase information is lost and the on-sky probability density appears roughly uniform. Near the location of the original detection (and its opposite point on the sky), the distribution pinches down considerably. This pinching occurs because all allowable orbits must pass through the location (and its opposite node) where the discovery was made. 

We also ran an orbit fit where we allowed the eccentricity of the orbit to be a free parameter. We include a prior proportional to $1/\sqrt{e}$, which incorporates our prior belief that eccentricity should be low due to tidal damping. We also placed an upper limit on $e < 0.25$---similar to Namaka's excited eccentricity \citep{ragozzine2009orbits,proudfoot2024bpm3}---and a lower limit on the periapse distance $q>4100$ km to prevent intersection with the rings. In this case, we found $a = 5910^{+702}_{-472}$ km, or orbit periods of $P_{orb} = 3.7^{+0.7}_{-0.4}$ days. We show the on-sky probability density 90 days after detections in Figure \ref{fig:ecc} to compare with the bottom panel of Figure \ref{fig:evolution}. The most immediate effect is a broadening of the probability distribution, although its highest density regions remain $\sim$unmoved. 


\begin{figure*}[t]
    \centering
    \includegraphics[width=\textwidth]{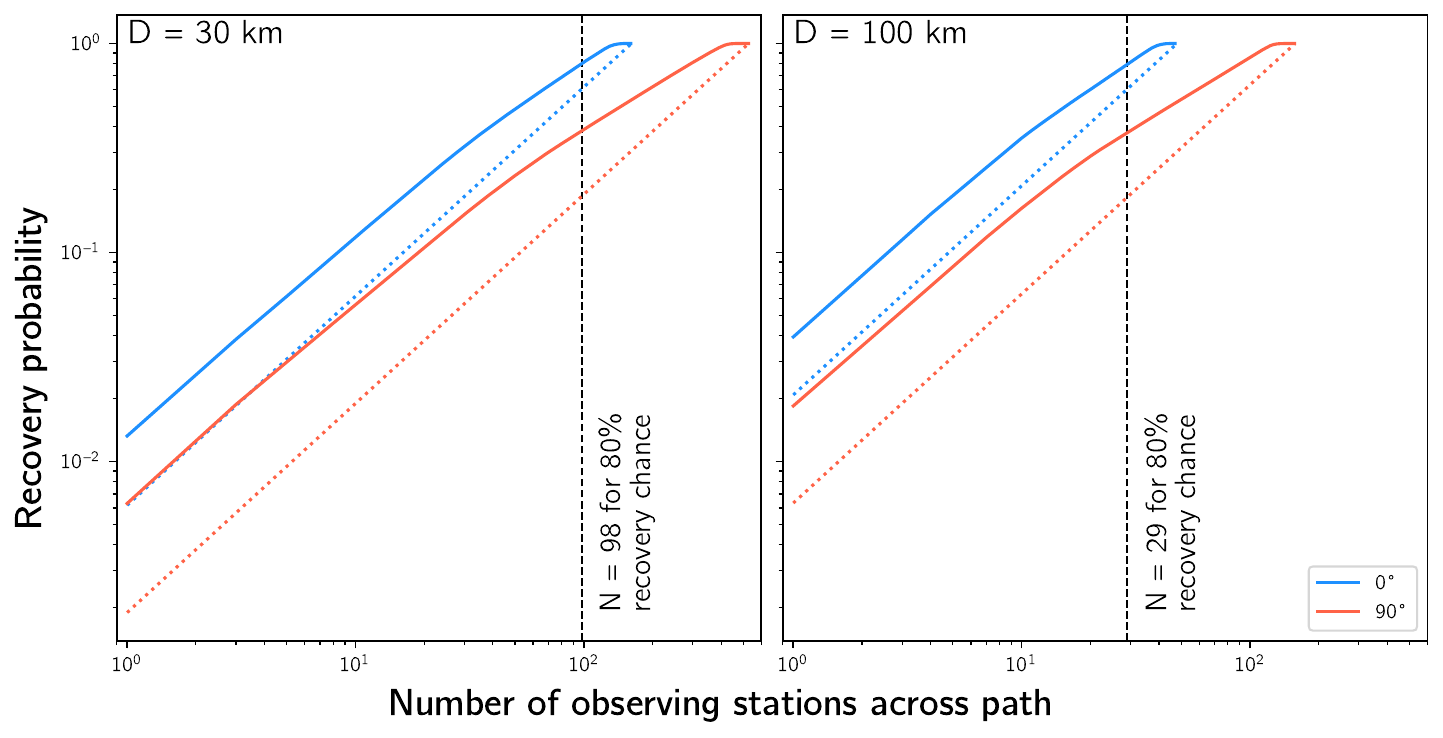}
    \caption{Probability of recovering the new satellite as a function of number of observing stations spaced at minimum a satellite diameter ($D$) apart. Solid lines show the probability when stations are placed in regions of highest probability first (in descending order) while dotted lines show the probability with random site selection (although with at least $D$ km apart). For $D = $ 30 km---the lower limit on the satellite's size---a campaign must have at least 98 observing stations to have a $>$ 80\% chance of recovery, assuming ideal station placement and perfect recovery if in the shadow path. }
    \label{fig:recovery}
\end{figure*}

Based on these orbit fits, we can also locate any resonances associated with the satellite's mean motion. Interestingly, we find that the (inner) 5:3 mean motion resonance with the new satellite is located at $4153^{+364}_{-232}$ km, close to the outer ring located at $4096\pm10$ km \citep{proudfoot2025constraints}. The 5:3 resonance is a strong, second-order resonance and could play a further role in Q1R's dynamics. The arcs seen in Q1R remain mysterious and could plausibly be created by confinement within such a resonance. The satellite's (inner) 7:2 resonance is also near Q2R, while the satellite is also quite close to Weywot's (inner) 7:2 resonance \citep[][]{Nolthenius_2025}. Unfortunately, these are quite weak fifth-order resonances that seem unlikely to play a significant dynamical role. At this time, we caution against over-interpreting possible resonance placement as the uncertainties are substantial. Future improvements in orbit fits will allow these apparent commensurabilities to be studied in greater detail.

\section{Recovery With Stellar Occultations}
\label{sec:occultations}

We can use the predicted probability distribution of on-sky positions to explore the possibility of targeted recovery of the satellite with stellar occultations. In this section, we assume that observing stations can be arbitrarily placed across the satellite's shadow path and that any observing station within it will detect the satellite. As seen in Figure \ref{fig:evolution}, the on-sky probability distribution of the satellite's position eventually converges to a stationary distribution that no longer evolves in time. Hence, we take the probability distribution at 180 days after discovery and use it as a template for exploring stellar occultation campaigns. 

Each possible stellar occultation traces out a $\sim$straight line on the sky with a position angle $\theta$. For simplicity, we define $\theta$ relative to the ring plane position angle, where $\theta = 0\degr$ (90$\degr$) is parallel (perpendicular) to the projected major axis of the rings. We then place $N$ observing stations across the occultation path with spacing of the satellite diameter, $D$. This allows us to determine the number of observing stations to guarantee a recovery of the satellite. For $D$ = 30 km, $N\sim150$ (500) for $\theta = 0\degr$ ($90\degr$). For $D$ = 100 km, $N\sim50$ (150) for $\theta = 0\degr$ ($90\degr$). 

With occultation chords crossing the probability distribution at some random angle, some observing stations will have higher detection probability than others. If observing stations are selected in order of lowest to highest probability of detection, a more efficient campaign can be carried out. For example, a 80\% detection probability can be achieved (with $D$ = 30 km) with 98 observing stations (compared to $\sim$150 for 100\% probability). In Figure \ref{fig:recovery}, we show the detection probability as a function of number of observing stations, $D$, and $\theta$. We show the improvement given by efficient placement (highest-to-lowest probability) and random placement. 

\begin{figure*}
    \centering
    \includegraphics[width=\textwidth]{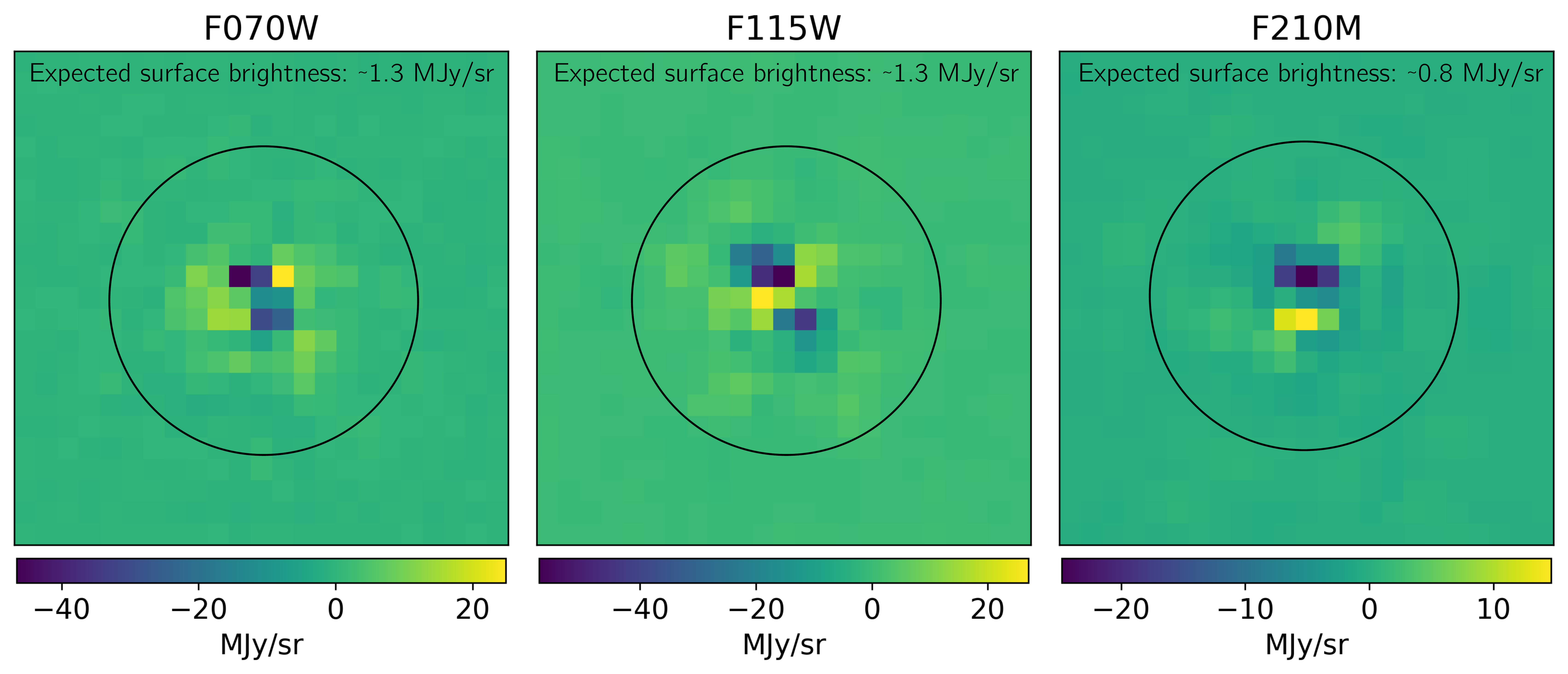}
    \caption{PSF fitting residuals of JWST images of Quaoar. Black circles show a 0.21$''$ radius (7 NIRCam pixels), within which the newly discovered satellite should be. The expected surface brightness of the satellite in the NIRCam images (assuming $V=28$ mag) is $\sim$1 MJy sr$^{-1}$. No clear detection is apparent, with PSF subtraction residuals dominating the region in which we expect the satellite to be. Without improvements to the NIRCam PSF model, or high quality empirical PSFs, recovery of the satellite will be $\sim$impossible.}
    \label{fig:images}
\end{figure*}

Given the large number of observing stations required for recovery---even in our idealized exploration---we expect that recovery of the satellite will be challenging. 
If small ($D$ = 30 km), recovery will take $\sim$hundreds of observers over several stellar occultations. Unfortunately, the late discovery of this small satellite---more than a decade after the first successful occultations of Quaoar---suggests that it is indeed likely to be small. 


\section{Possibility of Direct Imaging}
\label{sec:telescopes}
Since the discovery of Weywot in 2006 \citep{brown2007satellites}, there has been extensive imaging of the Quaoar system with the Hubble Space Telescope (HST). Despite HST's location in space, its relatively small aperture limits the detectability of small satellites around Quaoar, both in terms of angular resolution and signal-to-noise. As seen with past HST observations, Weywot's detectability is fairly limited when within 0.2$''$ of Quaoar \citep{fraser2010quaoar}, similar to the maximum separation of the new moon. Looking to the SNR achievable with HST, both current and defunct instruments produce SNR $\lesssim$5 (with a $V=28$ source), if exposing for an entire HST orbit. Hence, the satellite is likely undetectable with HST (or in any archival images). 

We evaluated the potential to detect the newly discovered satellite in JWST/NIRCam images using a point spread function (PSF) fitting technique. Specifically, we examined the images of the Quaoar system obtained in program \#6064 (PI: Souza-Feliciano) with the F070W, F115W, and F210M filters. These filters are used for the short-wavelength channel of NIRCam, which has a pixel scale of 0.031$''$/pixel \citep{Rieke2023}; the F070W filter, in particular, has the smallest achievable PSF full-width at half-maximum (FWHM) of any JWST observing mode.

The calibrated images (\dataset[available at doi: 10.17909/nrpb-6h26]{http://dx.doi.org/10.17909/nrpb-6h26}) were retrieved from the Mikulski Archive for Space Telescopes (MAST). Distorted, detector-sampled PSF models in each filter were generated using the STPSF tool \citep{perrin2012simulating,perrin2014updated}, with the approximate position of Quaoar on the detector specified to improve the accuracy of the distortion. The PSFs were fit using a 9-parameter model: the (x, y) positions of the Quaoar and Weywot PSF centroids, the flux scaling for each PSF, a PSF broadening factor for each PSF, and the background level. The PSF broadening factor accounts for any brighter-fatter effect (BFE) issues, telescope jitter, and deviations from a point source (specifically in the case of Quaoar). The results of the fits are presented in Figure \ref{fig:images}. Even accounting for detector-location-specific distortion, sub-pixel PSF positions, and PSF broadening, the residuals are too large to identify the new satellite, which is estimated to have a surface brightness of $\sim$1 MJy/sr within $\sim7$ pixels of the Quaoar centroid.

Even though the JWST exposure time calculator (ETC) estimates a SNR of $\sim$20 in the F210M filter (for $V\sim28$ mag), PSF residuals make satellite detection $\sim$impossible, except \textit{possibly} at maximum separation. With the exposure times required for detection ($\sim$hours), smearing due to the satellite's motion may begin to further degrade detectability. Longer exposures in the shortest filters (smallest FWHM) \textit{may} be able to retrieve the satellite at maximum separation, but without any phase information, retrieval will have to happen blindly, requiring large investments of telescope time.

With questionable direct detectability using current-generation telescopes, future telescopes may be required to directly observe the newly discovered satellite. Thirty-meter class telescopes such as the Extremely Large Telescope or Giant Magellan Telescope may provide the means to directly image Quaoar, its ring system, and its small satellite system using adaptive optics. Large space-based telescopes like Habitable Worlds Observatory may also provide similar capabilities. 

\section{Discussion}
\label{sec:discussion}

Once recovered, how many detections are needed to fully characterize the orbit of this small satellite? Operating under the assumption of a Keplerian orbit, the satellite's orbital motion is determined by seven parameters: the total system mass and six Keplerian orbital elements. Assuming the mass of the satellite is negligible, Quaoar's mass has already been fairly well-measured by modeling of Weywot's orbit (with Weywot making only a $\sim$0.1\% contribution to the total mass). Technically, only three detections are \added{theoretically} needed for a unique orbit solution; however, in practice, typically five detections are \added{ideal} \citep{grundy200842355}. \added{These additional detections help to eliminate mirror orbit solutions present when the system is only seen at a small range of viewing geometries, as well as eliminating possible period aliases.} Thankfully, under a realistic set of assumptions (as quantified by priors like those we use above), even a single additional detection can provide significant constraints on the satellite's orbit, allowing for better occultation predictions.


The discovery of this small satellite near Quaoar's rings provides evidence that Quaoar's rings and satellites formed from an initially broad and dense collisional disk around Quaoar, as suggested by \citet{sicardy2025rings}. Both satellites likely accreted from this disk while the known rings are presumably remnants of that disk, located in small niches where dynamical interactions stir the debris vigorously enough to prevent accretion. As pointed out by \citet{morgado2023dense}, the Roche limit may actually be much further from Quaoar due to more elastic collisions at lower temperatures in Quaoar's rings (30-40 K). However, this does not explain the confinement of today's rings, especially since the excited dynamical environment will cause rapid viscous spreading. Spin-orbit or mean motion resonances likely play a role in confining the rings, but tidal migration of Weywot will cause outward migration of both types of resonances, suggesting the ring system has significantly evolved since its formation. Perhaps the diffuse rings could effectively couple to various resonances and slowly migrate to their current locations. 

Interestingly, as seen in Figure \ref{fig:laplace}, the rings extend out to the radius beyond which the local Laplace plane begins to warp---assuming a reasonable density for Weywot. Beyond $\sim$4000--5000 km, warping of the local Laplace plane could introduce significant warps to a ring/disk, which would naturally excite the collisional activity of the rings and accelerate erosion/viscous spreading. While this does not explain the lack of accretion within the more stable inner regions, it may provide a limit on the radial extent of the rings. Although possibly coincidental, this further suggests an interlinked relationship between the rings and Quaoar's satellite system. 

Considering how Quaoar's ring-satellite system co-formed/evolved, one compelling idea is that the new satellite is the product of a now-accreted ring. Given a diameter of 30 km and a density of 800 kg m$^{-3}$, the mass would be equivalent to a $\sim$400 km wide ring at a radius of 5800 km, assuming a surface density of 750 kg m$^{-2}$ (similar to dense portions of Saturn's rings). The current rings are likely to be far less dense than this, but they may have been denser in the past. In contrast, Weywot would have to be made from a far denser ring/disk, as Weywot is $\sim$100$\times$ more massive than the new satellite. Alternatively, Weywot could be an intact fragment of an impactor early in Quaoar's formation \citep[e.g.,][]{arakawa2019early}. Such collisions may produce extensive disks of debris and small fragments, which may explain the new satellite and ring system. More detailed smoothed particle hydrodynamics simulations, combined with observations of the rings and characterization of small moons, will provide detailed insight into the origin of the Quaoar system.

\section{Conclusions}
\label{sec:conclusions}
In this letter, we constrain the orbital characteristics of a newly discovered satellite around Quaoar. Our main conclusions are:

\begin{enumerate}
    \item Based on the discovery observations, and assuming a circular orbit, we estimate that the newly discovered satellite is on a $3.6^{+0.5}_{-0.3}$ day orbit, which places Quaoar's outer ring near its 5:3 mean motion resonance.
    \item Stellar occultations are able to recover the satellite, but would require substantial community participation. We estimate that hundreds of observing stations would be required. Once recovered a few times, occultation predictions will vastly improve, providing ample opportunity for physical characterization. We call for broad participation in occultation campaigns to recover the new satellite, discover additional small satellites, and map the rings of Quaoar. Quaoar is favorably placed in the Scutum Star Cloud for another ~10 years, providing the best opportunities for occultations during its 286 year orbit.
    \item Current ground- and space-based telescopes will struggle to detect the newly discovered satellite due to both its faintness ($\sim$9-10 mag fainter than Quaoar) and small angular separation from Quaoar ($\lesssim0.2''$). Our examination of JWST NIRCam images of the Quaoar system show no compelling detection of the satellite. Direct imaging with current facilities will require considerable investment of telescope time to blindly reacquire the phase of the satellite, if it is indeed detectable. Future generation telescopes, however, will likely be able to easily observe it.
    \item The discovery provides evidence for an initially broad disk around Quaoar. Examining the formation and history of the moon-disk system will provide a detailed view into the formation of TNOs. We encourage sophisticated tidal, hydrodynamical, and collisional modeling of the Quaoar system.
\end{enumerate}

\begin{acknowledgments}

We acknowledge the Lucky Star collaboration for their work to continually update the ephemerides of TNOs.

This work is based in part on observations made with the NASA/ESA/CSA James Webb Space Telescope. The data were obtained from the Mikulski Archive for Space Telescopes at the Space Telescope Science Institute, which is operated by the Association of Universities for Research in Astronomy, Inc., under NASA contract NAS 5-03127 for JWST. These observations are associated with programs \#6064.

BP and FLR acknowledge the support of the UCF Preeminent Postdoctoral Program (P$^3$). EFV acknowledges support from the Space Research Initiative (SRI). RN acknowledges partial support from Cabrillo College.

\end{acknowledgments}

\begin{contribution}

BP provided orbit characterization, exploration of stellar occultations, writing the manuscript, and submitting it. RN was the satellite discoverer and provided details of its discovery circumstances. BJH provided PSF fitting for JWST data. ACDS was the PI of the JWST imaging program. FLR, CC, WG, EFV, and all other coauthors provided feedback, editing, and writing support. 

\end{contribution}

\bibliography{all}{}
\bibliographystyle{aasjournalv7}

\end{document}